\documentclass[prl,aps,twocolumn,footinbib,superscriptaddress,floatfix,bibliography]{revtex4-1}
\usepackage{pdfpages}
\usepackage{CJK}
\usepackage[colorlinks,bookmarks=false,citecolor=blue,linkcolor=red,urlcolor=blue]{hyperref}
\usepackage{color} 
\usepackage{graphicx}
\usepackage{float}
\usepackage{multirow}
\usepackage{amsmath}
\usepackage{amsfonts}
		
\usepackage[caption=false,subrefformat=parens,labelformat=parens]{subfig}

\newcommand{\curl}{{\nabla_\perp}}
\newcommand{\CCO}{Ca$_{10}$Cr$_7$O$_{28}$}
\newcommand{\TTO}{Tb$_2$Ti$_2$O$_7$}
\newcommand{\NZO}{Nd$_2$Zr$_2$O$_7$}

\makeindex

\makeatletter
\AtBeginDocument{\let\LS@rot\@undefined}
\makeatother
\begin{document}
\graphicspath{{./figs/}}
\begin{CJK*}{UTF8}{gbsn} 

\title{Half moons are pinch points with dispersion}

\author{Han Yan (闫寒)}
\email{han.yan@oist.jp}
\author{Rico Pohle}
\email{rico.pohle@oist.jp}
\author{Nic Shannon}
\email{nic.shannon@oist.jp}
\affiliation{Okinawa Institute of Science and Technology Graduate University, Onna-son, 
Okinawa 904-0412, Japan}

\date{\today}

\begin{abstract}

``Pinch points,'' singular features observed in (quasi-)elastic neutron scattering, 
are a widely discussed hallmark of spin liquids with an emergent gauge symmetry.
Much less attention has been paid to ``half moons,'' distinctive 
crescent patterns at finite energy, which have been observed in experiments on 
a number of pyrochlore magnets, and in a wide range of model calculations.
Here we unify these two phenomena within a single framework, paying 
particular attention to the case of ordered, or field-saturated states, where 
pinch points and half moons can be found in bands of excitations above a gap.
We find  that half moons are nothing other than pinch points inscribed 
on a dispersing band.
%
Molecular dynamics simulations of the kagome lattice antiferromagnet are used 
to explore how these bands evolve into the ground state and excitations 
of a classical spin liquid.
We explicitly demonstrate that this theory can reproduce the pinch 
points and half moons observed in \NZO.
\\
%
\end{abstract}
\maketitle
\end{CJK*}



{\it Introduction. } 
A central challenge in the study of frustrated magnets is to identify features which 
can be used to distinguish between different types of magnetic states, in the absence 
of conventional long--range magnetic order \cite{Balents2010,Savary2016}.
In this context, any robust feature observed in more than one system 
is of potential interest as a witness to the underlying physics.
A prime example is provided by ``pinch points'' --- singular, bow-tie-like 
motifs in the spin structure factor, characteristic of ``Coulombic'' phases with an 
emergent gauge 
symmetry~\cite{moessner98-PRB58,huse03,henley10,Benton2016,prem-arXiv}. 
Pinch points have famously been observed in neutron-scattering experiments 
on spin ice \cite{Fennell2009}, a wide range of other pyrochlore magnets \cite{fennell12,Sibille2018}, 
and in simulations of, e.g., kagome-lattice antiferromagnets \cite{chalker92,Zhitomirsky2008, Taillefumier2014}.


Another characteristic feature, often observed in parallel with pinch points, are the split rings 
of scattering found at finite energy in Tb$_2$Ti$_2$O$_7$ \cite{Guitteny2013,Fennell2014}; 
in the excitations of the ``proximate'' spin-liquid \NZO\ \cite{Petit2016,Lhotel2017}, 
and in numerical simulations of a wide range of frustrated magnets, where they have been 
described as ``excitation rings'' \cite{Robert2008,Taillefumier2014}, ``spherical surfaces'' \cite{Rau2016}, and  
``half moons'' \cite{Udagawa2016, Mizoguchi2017}.
However, despite being documented a decade ago \cite{Robert2008}, the connection 
between pinch points found at low energy, and the half moons observed at higher energy, 
remains obscure.


\begin{figure}[t]
	\centering
		\subfloat[\label{fig:1_1_1}]{\includegraphics[height=5.5cm]{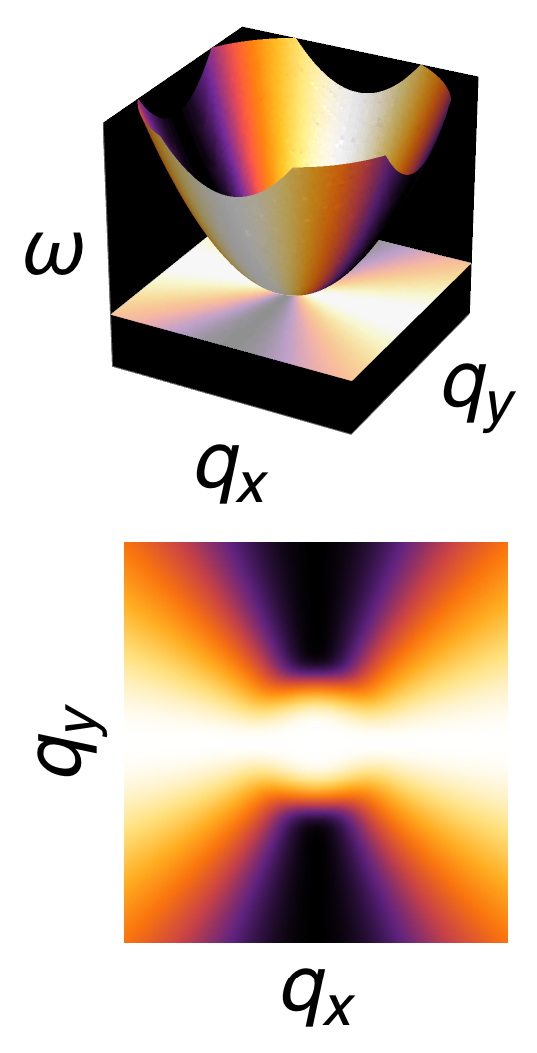}}
		\subfloat[\label{fig:1_1_2}]{\includegraphics[height=5.5cm]{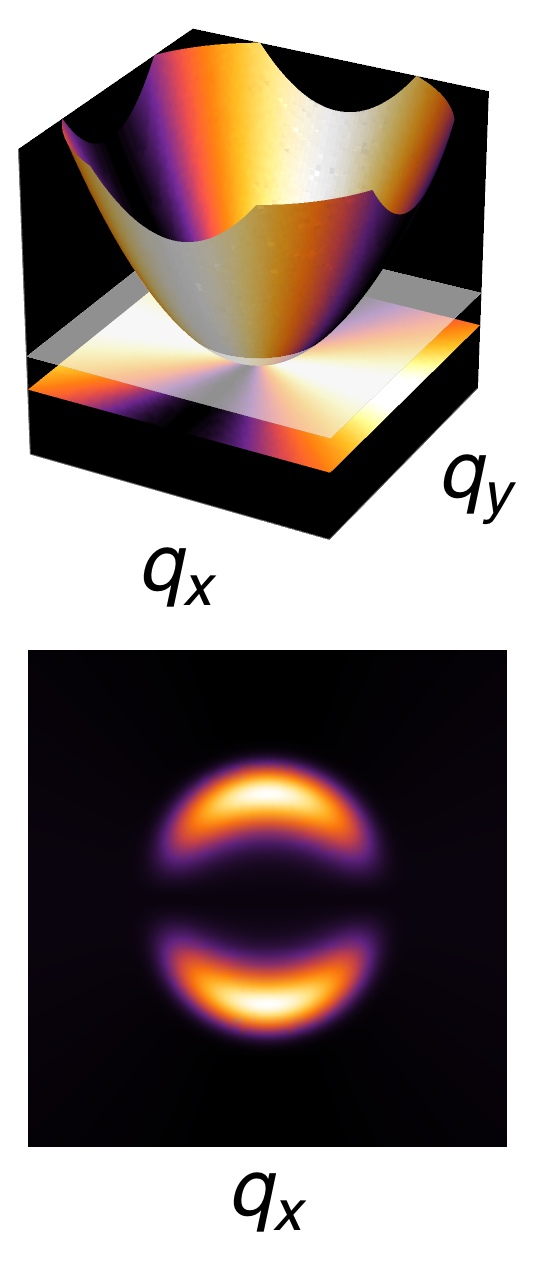}}
		\subfloat[\label{fig:1_1_3}]{\includegraphics[height=5.5cm]{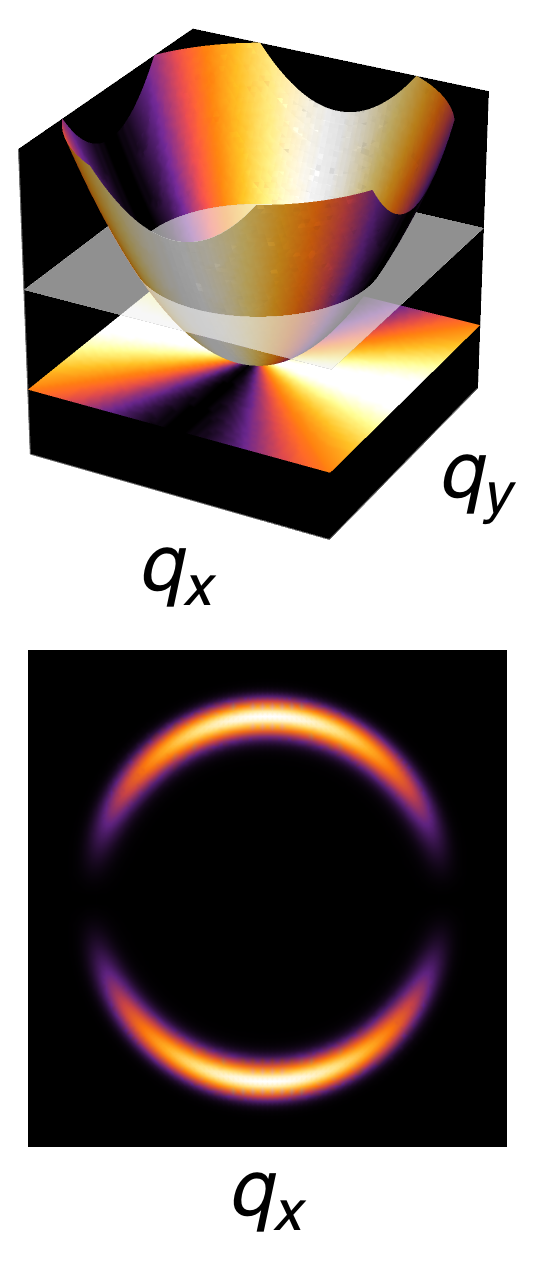}}
		\subfloat{\raisebox{1.25cm}{\includegraphics[height=3.0cm]{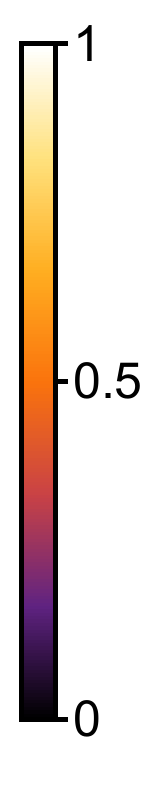}}}
	\caption{
		Illustration of connection between pinch points and half moons, as found in the
		saturated phase of 
		the Heisenberg antiferromagnet (HAF) on a kagome lattice, in applied magnetic field.
		Upper panels: flat and dispersing bands of spin excitations, 
		showing cross sections at fixed energy (white plane).  
		Lower panels: corresponding prediction for the dynamical structure factor 
		at fixed energy.
		Pinch-point singularities are encoded in both the flat, and the dispersing band, 
		where they appear as half-moon features.
		Results have been calculated within a continuum field theory, described below, 
		and convoluted with a Gaussian envelope, to mimic the effect of 
		finite energy resolution.
		}
\label{fig:Fig_1_Zoom_in_Crescents}
\end{figure}


In this Rapid Communication, we establish a unified theory of pinch points and half moons, 
considering the simplest model which exhibits both features in its dynamical structure factor --- 
the Heisenberg antiferromagnet (HAF) on a kagome lattice, with magnetization saturated 
by applied magnetic field.
Introducing a description in terms of continuum fields, we show that pinch points 
and half moons arise from the divergence-free and curl-free components of the 
same, emergent, magnetic field.  
Regular pinch points arise from the divergence-free condition, and the associated 
spin excitations form a flat band.
In the case of the curl-free component, excitations form a dispersing band, 
while the intensity of the associated scattering is modulated in the same way  
as for a (rotated) pinch point.
The combination of these two effects leads to characteristic half-moon features in correlations 
at fixed energy --- a phenomenology summarized in Fig.~\ref{fig:Fig_1_Zoom_in_Crescents}.  


We further use molecular dynamics (MD) simulation to explore the fate of pinch-points and 
half-moons in the absence of magnetic field.
We find that the pinch-points and half-moons found in the saturated state
evolve smoothly into those observed for \mbox{$H \to 0$}.
Finally, we discuss the relevance of these results to real materials, using our theory 
to develop explicit predictions for the pinch points and half moons observed in \NZO, 
in excellent agreement with experiment.

\begin{figure}[t]
	\centering
	\subfloat{\includegraphics[width=2.5cm]{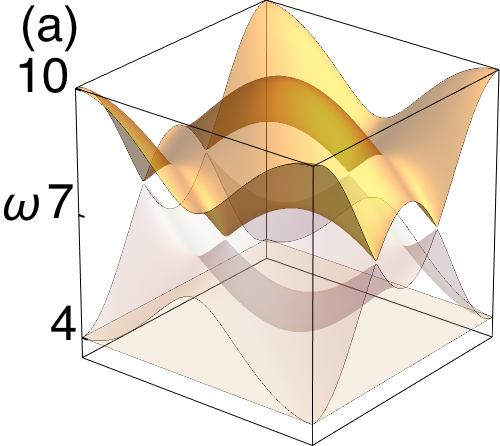}\label{fig:2-1-top-band}}\hfil
	\subfloat{\includegraphics[width=3.15cm]{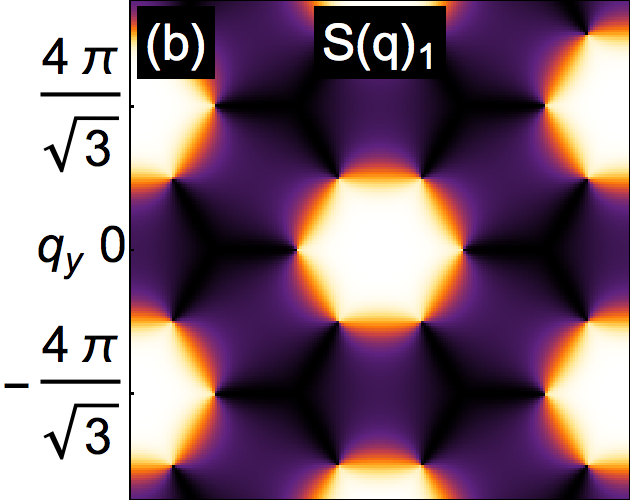}\label{fig:2-2-top-intensity}}\ 
	\subfloat{\includegraphics[width=2.5cm]{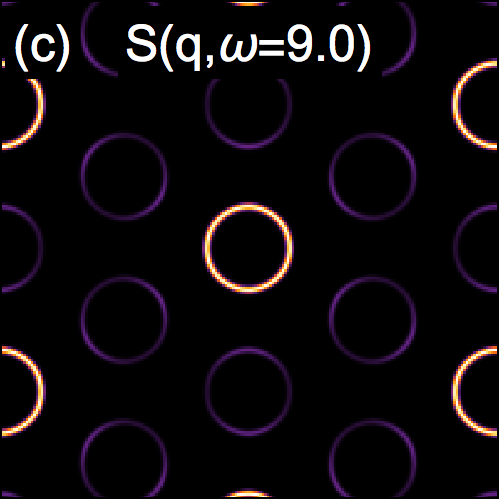}\label{fig:2-3-top-cut}} \\
	\subfloat{\includegraphics[width=2.5cm]{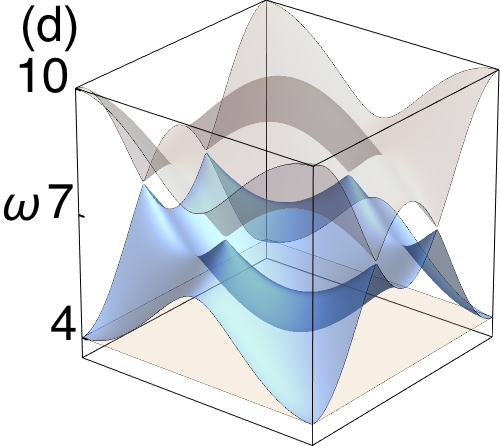}\label{fig:2-4-mid-band}}\hfil
	\subfloat{\includegraphics[width=3.15cm]{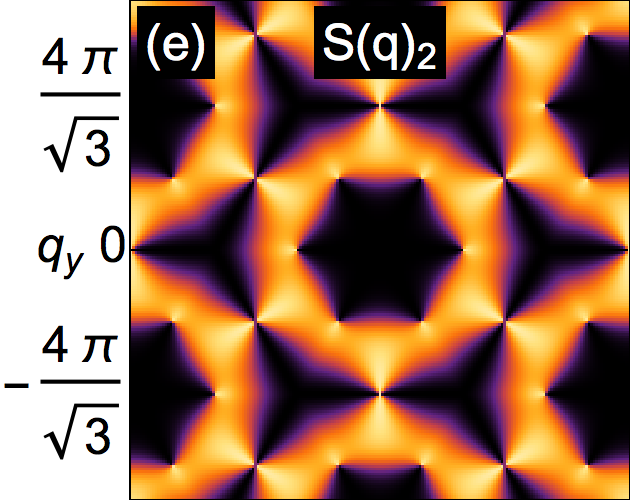}\label{fig:2-5-mid-intensity}}\ 
	\subfloat{\includegraphics[width=2.5cm]{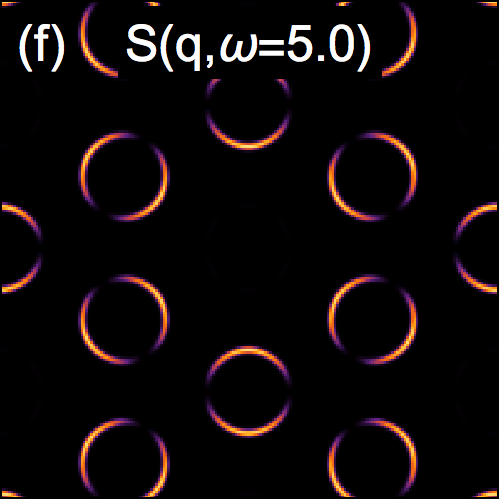}\label{fig:2-6-mid-cut}} \\
	\subfloat{\raisebox{0.325cm}
		{\includegraphics[width=2.5cm]{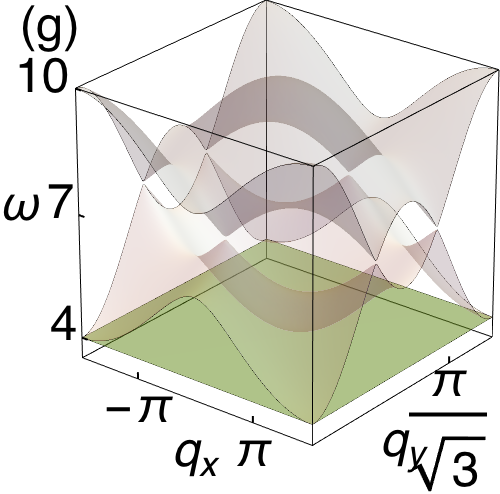}\label{fig:2-7-bot-band}}}\hfil
	\subfloat{\includegraphics[width=3.15cm]{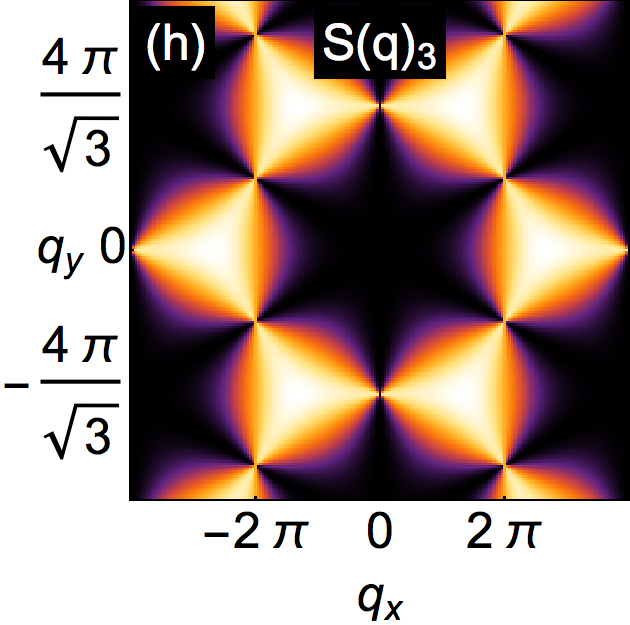}\label{fig:2-8-bot-intensity}}\ 
	\subfloat{\includegraphics[width=2.5cm]{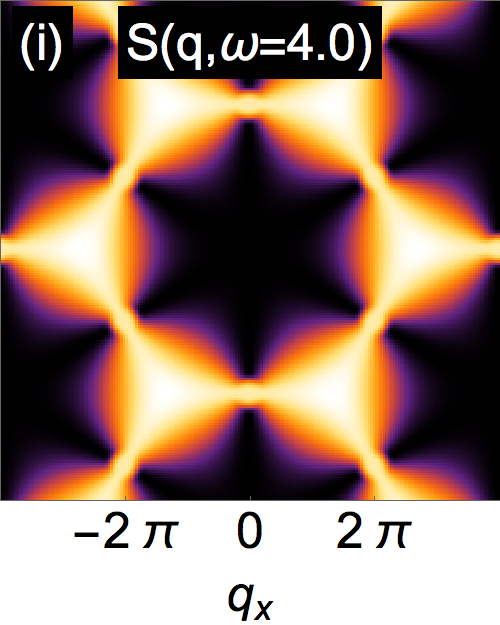}\label{fig:2-9-bot-cut}}\\
	\hspace{3cm}		\subfloat{\includegraphics[width=2.cm]{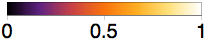}}
	\caption{
	 	Spin-wave excitations of the  Heisenberg antiferromagnet (HAF)  on a kagome lattice 
		in high magnetic field, showing associated pinch-point and half-moon features.
		(a)~Dispersion of upper band of excitations for \mbox{$7 \leq \omega \leq 10$}.   
		(b)~Contribution to equal-time structure factor, $S({\bf q})$, 
		coming from integration over the upper band of excitations.  	
		(c)~Dynamical structure factor $S({\bf q},\omega)$ evaluated at 
		$\omega=9.0$, showing rings of scattering.
		\mbox{(d)-(f)}~Corresponding results for the middle band of excitations, 
		showing half-moon features at $\omega=5.0$.
		\mbox{(g)-(i)}~Corresponding results for the flat band of excitations, 
		showing pinch-point features at $\omega=4.0$.
		All results were obtained within linear spin wave (LSW) theory for Eq.~\eqref{eq:H}, 
		with $J=1$ and $H=10$.}
	\label{fig:2_Plus_Limit}
\end{figure}

{\it The model. } 
We take as a starting point the HAF on a kagome lattice
\begin{equation}
\label{eq:H}
\mathcal{H}
= J \sum_{\langle i,j\rangle } {\bf S}_i \cdot {\bf S}_j 
 - H \sum_i S_i^z
\end{equation}
and consider first the case $H > 6J$, where the ground 
state has been saturated by magnetic field \cite{zhitomirsky02}.
In this case, one-magnon excitations are exact eigenstates, and can be 
calculated without approximation using linear spin wave (LSW) theory.
For \mbox{$H = 10J$}, this leads to the results shown in Fig.~\ref{fig:2_Plus_Limit}.
There are three inequivalent bands of excitations; a flat band at \mbox{$\omega = 4J$}, 
and two dispersing bands spanning energies \mbox{$4J \leq \omega \leq 10 J$}.
The flat band encodes pinch points, clearly seen in the dynamical structure factor 
$S( {\bf q}, \omega)$ for \mbox{$\omega = 4J$} [Fig.~\ref{fig:2-9-bot-cut}].  
Meanwhile, dispersing bands exhibit half-moon features, which ``grow out''
from those zone centers where pinch points are found at low energy 
[Fig.~\ref{fig:2-6-mid-cut}].  
The question is, how are these features connected ?


We can answer this question by introducing a continuum field theory 
description of the magnetic excitations.
We consider only the spin components
perpendicular to the magnetization, $\nu = x,y$, and group these 
into two vector and two scalar fields 
\begin{eqnarray}
{\bf m}_\nu  = \sum\limits_{i=1}^3 S^\nu_i {\bf u}_i 
, \qquad
\xi_\nu  =  \sum_{i=1}^3 S^\nu_i  ,
\label{eq:fields}
\end{eqnarray}
where the sum runs over the sites of a primitive unit cell, 
and the unit vectors 
\begin{equation}
\label{eqn:u-vec}
{\bf u}_1=(0,1),\quad {\bf u}_2=(-\sqrt{3}/2,1/2),\quad {\bf u}_3=(\sqrt{3}/2,1/2)  .
\end{equation}
It follows that the fields $\xi_\nu$ and ${\bf m}_\nu$ [Eq.~(\ref{eq:fields})] 
transform with the $A_1$ and $E$ irreps of the primitive unit cell, respectively \cite{Essafi2017}.
%
We can further separate the vector fields
${\bf m}_\nu$ into curl-full and divergence-full components by Helmholtz-Hodge 
decomposition \cite{arfken-book}:
\begin{equation}
{\bf m}_\nu 
= {\bf m}^\text{curl}_\nu + {\bf m}^\text{div}_\nu 
 ,\quad
{\bf m}^\text{curl}_\nu = \curl a_\nu
 , \quad
{\bf m}^\text{div}_\nu = -\nabla \phi_\nu, 
\label{eqn:field-decomp}
\end{equation}
where $a_\nu$ and $\phi_\nu$ are two scalar potentials and, in two dimensions, 
the curl and divergence are defined through
%
\begin{eqnarray}
\nabla_{\perp} = (-\partial_y,\partial_x) 
 , \qquad
\nabla = (\partial_x,\partial_y) .
\end{eqnarray}


This decomposition of the fields has much in common with the 
``moment fragmentation'' explored in the context of pyrochlore 
magnets \cite{Brooks-Bartlett2014,benton16-PRB94}, 
and the fact that ${\bf m}^\text{curl}_\nu$ obeys a zero-divergence condition 
naturally motivates the introduction of a $U(1)$ gauge field \cite{henley05,henley10}.
%
%
However within the high-field saturated state, all excitations are gapped,
implying that any emergent gauge symmetry has 
been lifted by a mass term in the effective 
action
\footnote{
See Supplemental Material for a more extended discussion,
, which includes 
Refs.~\cite{hermele04-PRB69,benton12-PRB86,lee12,kogut79}.}.
None the less, we can view the zero-divergence condition 
as a witness to a proximate gauge symmetry, which can be restored 
when the gap closes.
We return to this point in the context of simulations, below.


{\it Origin of pinch points.} 
Following \cite{benton16-PRB94}, we can now analyze spin-dynamics 
through the Heisenberg equations of motion (EoM), an approach which is expected to 
be exact for the one-magnon band at zero temperature.
%
Equation~(\ref{eqn:field-decomp}) leads to a remarkable simplification in the EoM, 
with ${\bf m}^\text{curl}_\nu$ decoupling from other fields entirely, to give 
\begin{eqnarray}
\partial_t a_x &=   \omega_0 a_y 
 , \qquad
\partial_t a_y =& - \omega_0 a_x  , 
\label{eqn:Bdfeom}
\end{eqnarray}
where here and below, we set $\hbar = 1$.  
This implies the existence of a flat band of excitations with energy 
\begin{eqnarray}
\omega^\text{curl}({\bf q}) = \omega_0 = H - 6J  .
\end{eqnarray}
The structure of correlations within this band are determined 
by the condition $\nabla \cdot {\bf m}^\text{curl}_\nu = 0$.
In momentum space, this implies ${\bf q} \cdot {\bf m}^\text{curl}_\nu = 0$, 
%
%
and the associated dynamical structure factor is given by
\begin{equation}
\langle m^\text{curl}_{\nu,\alpha} ({\bf q},\omega) 
m^\text{curl}_{\nu',\beta} ({\bf -q},\omega) \rangle  
\propto  
\delta_{\nu\nu'}\delta(\omega-\omega_0)
(1-{{q}_\alpha {q}_{\beta}}/{{\bf q}^2}) ,
\label{eqn:Bjpinchpointv2}
\end{equation}
%
%
where $\alpha, \beta = x,y$ are the spatial components
of the vector fields ${\bf m}^\text{curl}_\nu$ and ${\bf m}^\text{curl}_{\nu'}$
defined in Eq.~\eqref{eqn:field-decomp}.
It follows that the intensity of scattering has the familiar structure of a 
pinch point \cite{henley10}.
A parallel analysis, leading to pinch points on a flat band at finite energy, 
has been given by Benton \cite{benton16-PRB94} in the context of \NZO, a case which 
we return to below.


{\it Origin of half moons. }
We now turn to the dynamics of the field ${\bf m}^\text{div}_\nu$.   
In this case the relevant EoM are given by 
\begin{equation}
\partial_t \phi_x  = \left[\omega_0 +  D\nabla^2 \right] \phi_y  
, \qquad
\partial_t \phi_y = \left[-\omega_0 -  D\nabla^2 \right] \phi_x  
\label{eqn:jajbeom}
\end{equation}
yielding a band with quadratic dispersion 
\begin{equation}
\omega^\text{div}({\bf q}) = \omega_0 + D {\bf q}^2  , \qquad D = J/4  .
\end{equation}
The structure of correlations within this band are determined 
by the condition $\curl \cdot {\bf m}^\text{div}_\nu = 0$.    
This implies
\begin{eqnarray}
\tilde{\bf q} \cdot {\bf m}^\text{div}_\nu = 0
 , \qquad 
 \tilde{\bf q}  = (-q_y,q_x)    ,
\end{eqnarray}
and the corresponding dynamical structure factor is 
given by 
%
\begin{eqnarray} 
\langle  m^\text{div}_{\nu,\alpha} ({\bf q},\omega) 
m^\text{div}_{\nu',\beta} ({\bf -q},\omega) \rangle   
&\propto & 
\delta_{\nu\nu'}
\delta ( \omega - \omega_0 - D {\bf q}^2 ) \nonumber\\
&& \times (1-{{\tilde{q}}_\alpha {\tilde{q}}_{\beta}}/{{\bf q}^2})  .
\label{eqn:Brhopinchpoint}
\end{eqnarray}
These correlations also have the form of a pinch point, but this is now
imprinted on a dispersing band, and orientated perpendicular to the ``conventional'' 
pinch point in the flat band [Eq.~(\ref{eqn:Bjpinchpointv2})].   


The reason for the appearance of half moons in dynamical structure factors now 
becomes evident.    
In cuts taken at constant energy, the band of excitations associated with ${\bf m}^\text{div}_\nu$
appear as rings of scattering satisfying $\omega = \omega^\text{div}({\bf q})$, but with intensity 
which vanishes approaching a characteristic line in reciprocal space [cf. Eq.~(\ref{eqn:Brhopinchpoint})].
This converts a single ring into two, symmetrical, crescent features, aka~``half moons.''
In the case of the spin structure factor, $S({\bf q}, \omega)$, the orientation of the half-moons 
depends on which component of ${\bf m}^\text{div}_\nu$ is probed in a given Brillouin 
zone (BZ) (Fig.~\ref{fig:2-6-mid-cut}).    
However, within any given BZ, the half-moon feature is orthogonal to the 
pinch point in the accompanying flat band (Fig.~\ref{fig:2-9-bot-cut}).
This generic structure, of half moons (associated with a field satisfying a zero-curl 
condition), dispersing out of zone center where they meet a flat band (associated with 
a field satisfying a zero-divergence condition), is illustrated in Figs.~\ref{fig:1_1_1}-\ref{fig:1_1_3}.


The results of this field theory [Fig.~\ref{fig:Fig_1_Zoom_in_Crescents}] perfectly reproduce LSW 
calculations [Fig.~\ref{fig:2_Plus_Limit}], in the relevant 
long-wavelength limit.
Comparing the separate 
contribution of each spin-wave band  
to the equal-time structure factor $S({\bf q}) = \int d\omega S({\bf q},\omega)$, 
%
%
we again see that the pinch points of the dispersing band 
[Fig.~\ref{fig:2-5-mid-intensity}], are perpendicular to those of the flat band
[Fig.~\ref{fig:2-8-bot-intensity}], with the total spectral weight at each ${\bf q}$
satisfying a sum rule across the three bands.


\begin{figure}[t]
	\captionsetup[subfigure]{position=top}
	\centering			
	\subfloat{\includegraphics[height=3.8cm]{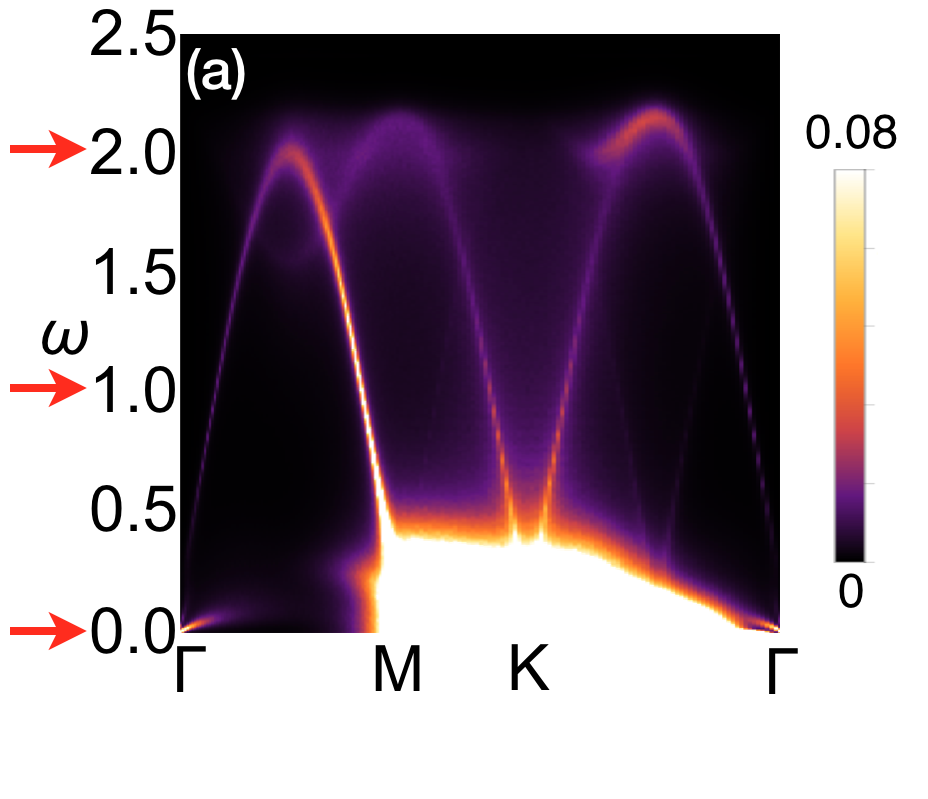}\label{fig:4_1}}
	\subfloat{\includegraphics[height=3.8cm]{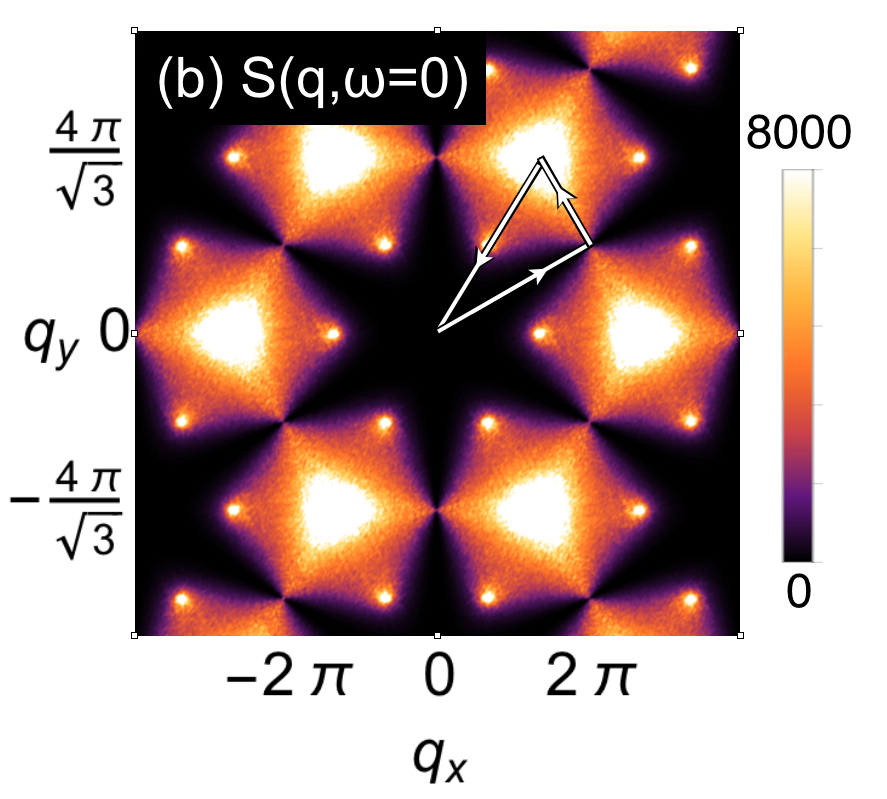}\label{fig:4_3}}\\
	\subfloat{\includegraphics[height=3.8cm]{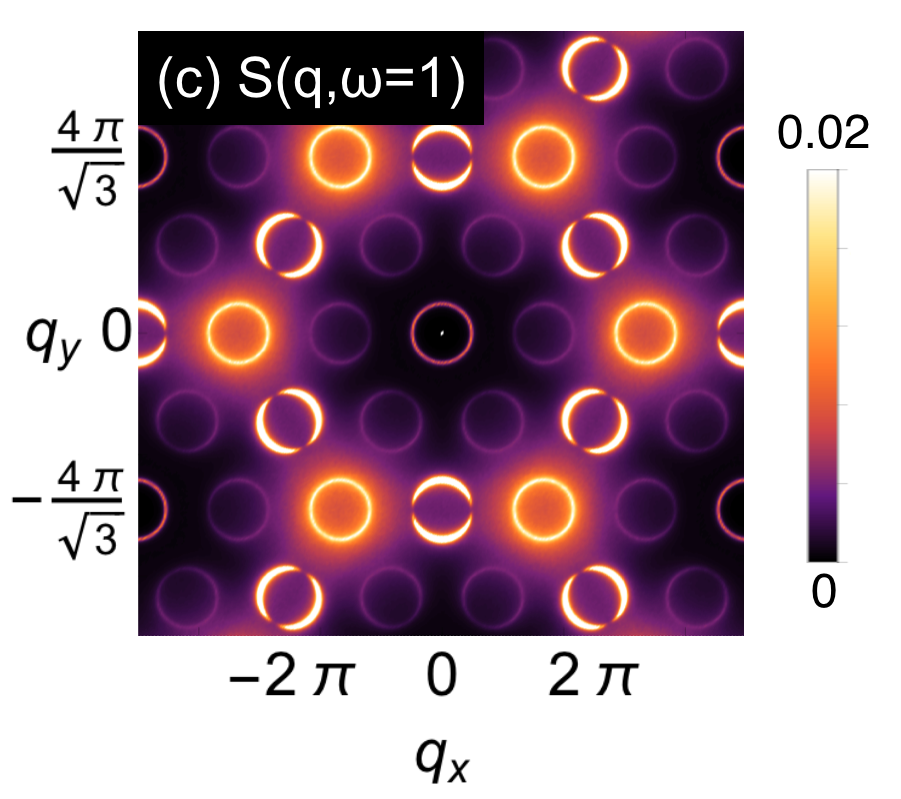}\label{fig:4_4}}\
	\subfloat{\includegraphics[height=3.8cm]{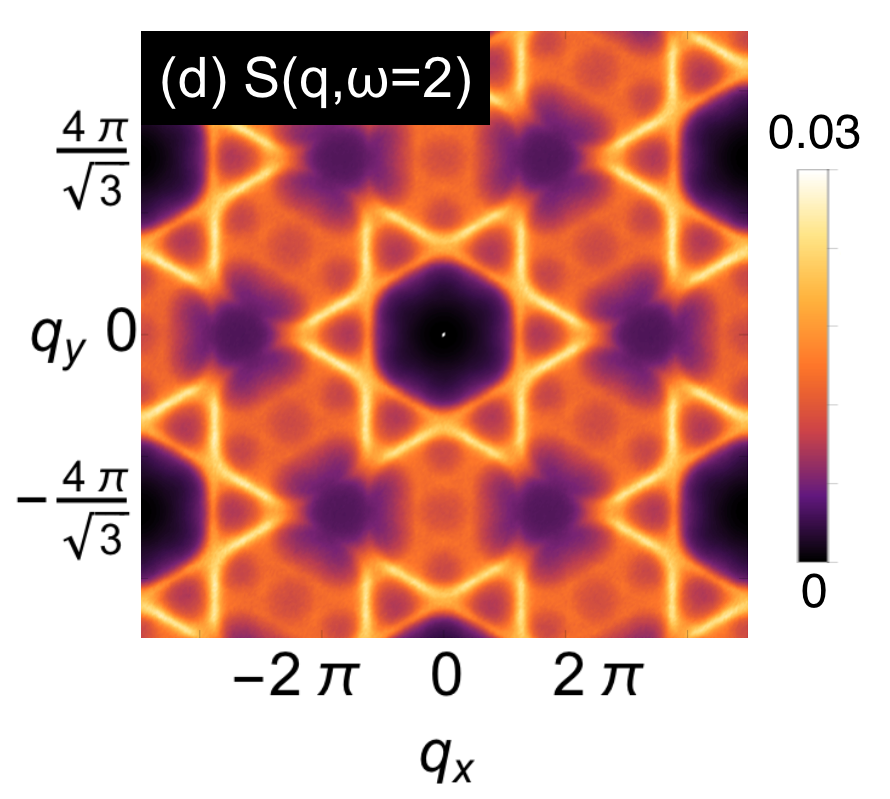}\label{fig:4_5}}\\
	\caption{
		Dynamical structure factor found in molecular dynamics (MD) simulations 
		of the kagome antiferromagnet in absence of magnetic field, showing persistence 
		of half-moon features.
		(a) Results for the dynamical structure factor $S({\bf q}, \omega)$ at $T=0.001J$, 
		on an irreducible wedge of the BZ.
		Energies used for cross sections at fixed $\omega$ are marked with red arrows.
		(b)-(d) Dynamical structure of HAF at $\omega=0,1,2$, 
		showing evolution of half-moons out of a flat band encoding pinch points at zero energy.
		}
	\label{fig:4}
\end{figure}


{\it  ``Half moons'' in the absence of magnetic field. }
So far, we have limited our discussion to gapped excitations about 
a field-saturated state.
Nonetheless, molecular-dynamics (MD) simulations 
have also revealed half-moon structures in the excitations of 
a (classical) spin-liquid on the kagome lattice in the absence of 
magnetic field \cite{Robert2008,Taillefumier2014}, 
and it is interesting to ask how pinch points and half moons 
evolve, once the gap to spin excitations closes.


In Fig.~\ref{fig:4} we show MD results in the absence of 
magnetic field ($H=0$).
We find a spin-liquid ground state, formed when the gap to 
the flat band closes \cite{zhitomirsky02} at the critical field $H_c = 6J$.
Half-moon features survive at finite energy [Fig.~\ref{fig:4_4}], with 
intensity perpendicular to the pinch points in the static structure factor [Fig.~\ref{fig:4_3}].
The bands carrying half moons evolve smoothly out of the excitations of the 
field-saturated state, and the structure of the half moons remains well-described 
by Eq.~\eqref{eqn:Brhopinchpoint}.   
These results are consistent with the restoration of an emergent $U(1)$ gauge symmetry 
at the field $H=H_c$.
And in this context, it is interesting to recall that the curl-free excitations, which give rise to half moons, 
reflect the field ${\bf m}^\text{div}$  [Eq.~(\ref{eqn:field-decomp})], coming from 
a net source of (emergent) magnet field, e.g. a magnetic monopole
\footnote{
See Supplemental Material 
for details of numerical simulations,
which include 
Refs.~\cite{Olive1986, Miyatake1986,Swendsen1986, Earl2005,Creutz1987,NumericalRecipes2007, OrdinaryDiffEquations1,FFTW05,arfken-book}.}
.


{\it Experimental realizations. }
The most straightforward experimental application of these ideas is to \NZO, a pyrochlore magnet
in which a Nd$^{3+}$ moment is  ``fragmented,'' so that a magnetically ordered ground state 
is accompanied by a ``dynamical spin liquid,'' with flat band encoding spin-ice-like 
pinch points above a finite energy gap \cite{Petit2016}.
At higher energies inelastic neutron scattering experiments 
perfectly capture the pattern of half moons on a quadratically dispersing band, 
evolving out of the pinch points of the flat band \mbox{(cf. Fig. S3 of \cite{Petit2016})}.
The microscopic structure of EoM for spin excitations in \NZO,  as well as parameters 
for a microscopic model, have already been discussed by Benton \cite{benton16-PRB94}.
What remains is to show that a field-theory of the type developed in this Rapid Communication 
can reproduce the scattering seen in experiment.

\begin{figure}[t]
	\centering
	\subfloat{\includegraphics[width=4cm]{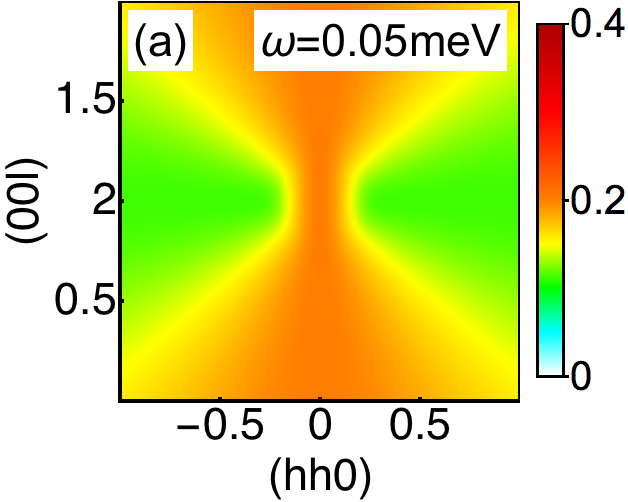}\label{fig:5-1}}\hfill
	\subfloat{\includegraphics[width=4cm]{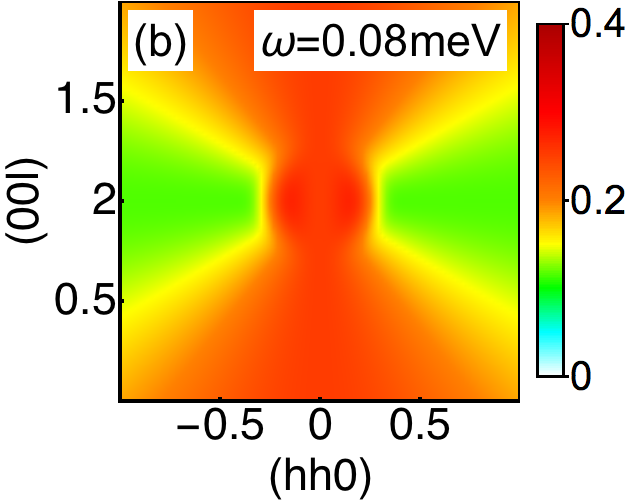}\label{fig:5-2}} \\
	\subfloat{\includegraphics[width=4cm]{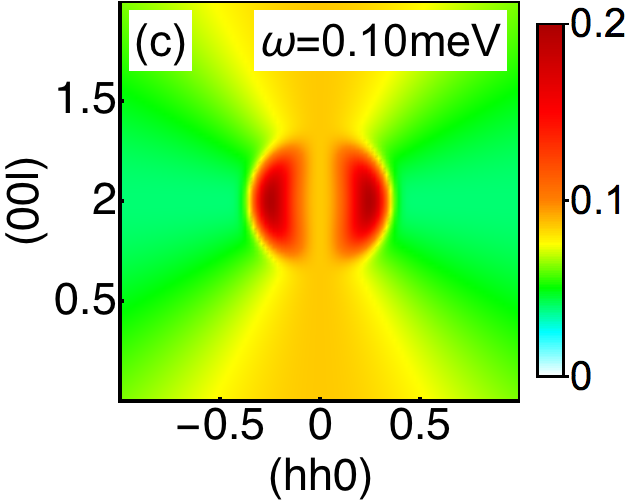}\label{fig:5-3}}\hfill
	\subfloat{\includegraphics[width=4cm]{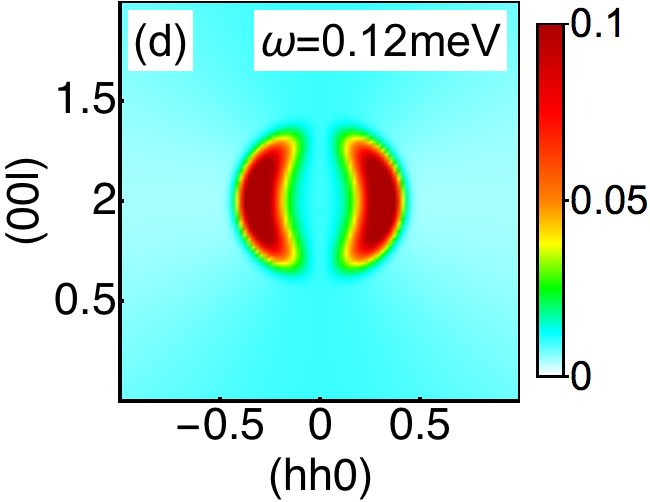}\label{fig:5-4}}
	\caption{
	         Predictions for the dynamical structure factors of the proximate spin liquid \NZO, 
	         calculated from a continuum field theory.
	         %
		The analysis has been parameterized from 
		experiment \cite{benton16-PRB94,Petit2016}, 
		leaving no adjustable parameters.
                 %
		%
		Approaching the zone center, the agreement with inelastic neutron scattering  
		is excellent (cf. Fig.~S3 of \cite{Petit2016}).}
	\label{fig:5_NZO}
\end{figure}

In Fig.~\ref{fig:5_NZO}, we present explicit field-theory predictions 
for \NZO, following the pattern developed above
\footnote{
See Supplemental Material 
for a more detailed derivation, which includes
Refs.~\cite{Petit2016,Benton2016}
}.
The field theory has been parameterized from the 
microscopic model of \cite{benton16-PRB94}, 
with overall intensity and experimental resolution   
determined from a fit to the elastic line in Fig.~4(a) of \cite{Petit2016}, 
leaving no adjustable parameters.
In this case our theory is tied to a mean-field approximation of the 
ground state, and is only exact in the limit where the gap
to spin excitations is much larger than the interactions between them.   
None the less the agreement with experiment 
is excellent, confirming that  half-moon features originate in a dispersing band of 
excitations satisfying a zero-curl condition.   
We note that behavior consistent with half-moon features dispersing out of a 
flat band has also been observed for \NZO\ in a magnetic field parallel to the 
[111] axis, where it is expected to realize a dynamic quantum kagome 
ice \cite{Lhotel2017}.


\TTO, another Pyrochlore oxide, comes with more complication \cite{Guitteny2013,Fennell2014,Ruminy2016-2,Ruminy2016}.
In \cite{Guitteny2013}, a dispersive band with half moons
is observed at very low energy scale from 0.0 to 0.3 meV,
attaching to the pinch-point band at zero energy.
Remarking on these, the authors of \cite{Guitteny2013} conjectured that
the half moons {\it ``could be an intrinsic feature of Coulomb phases,"}
which {\it ``will have to be confirmed in further theoretical studies."}
The results in this Rapid Communication suggest that half moons are indeed intimately 
related to the emergent gauge structure of Coulomb phases.
As there is no established microscopic model for \TTO, it is not possible  
to carry out the same kind of quantitative analysis as for \NZO.
Nonetheless, it might be interesting to apply a phenomenology of the 
form of Eq~\eqref{eqn:Brhopinchpoint} to inelastic neutron scattering data.


Lastly, we turn to \CCO, a recently-discovered material realising a 
spin-1/2 Heisenberg model on a bilayer breathing-kagome (BBK) lattice, 
which supports a quantum spin liquid ground state \cite{balz16-NatPhys12}.  
Here once again, inelastic neutron scattering carried out in magnetic field
reveals both flat and dispersing bands at finite energy, with evidence for 
pinch-point structure within the flat band \cite{balz17-PRB95}.
Both numerical simulations and spin-wave calculations  \cite{Pohle2017}, parameterized 
from experiment  \cite{balz17-PRB95}, reveal distinct half-moon features dispersing 
out of the pinch points of the flat band, consistent with the framework 
in this Rapid Communication.
We therefore anticipate that more detailed measurements of the dispersing 
bands of \CCO\ in high field will reveal half-moon features.
\\


{\it Summary and conclusions. }
In this Rapid Communication we have explored the connection between pinch points --- singular 
features in neutron scattering associated with both Coulombic spin liquids and flat bands 
of excitations in  frustrated magnets; and  half moons --- characteristic 
crescent features in inelastic scattering which are frequently found accompanying 
pinch points.
We find that half moons are nothing but a second set of pinch points with dispersion 
[Fig.~\ref{fig:Fig_1_Zoom_in_Crescents}].
Considering the specific example of the Heisenberg antiferromagnet on a kagome lattice, 
in applied magnetic field  (Fig.~\ref{fig:2_Plus_Limit}),  
we show that the connection between pinch points and half moons 
can be made explicit through a Helmholtz decomposition of the associated spin 
configurations, implying that they originate in the same, proximate, $U(1)$ gauge 
symmetry.
While our field-theoretic anaylsis is limited to excitations of a field-saturated state, 
we expect the same considerations to apply to classical spin liquids, and  
find evidence in MD simulation to reinforce this claim (Fig.~\ref{fig:4}).
Finally, we have discussed the application of these ideas to the frustrated magnets 
\TTO\ and \CCO, and demonstrate that they quantitatively reproduce the half moon 
features found in \NZO\ (Fig.~\ref{fig:5_NZO}).


We conclude with a few comments about interesting open problems.
Half moons have also been observed in models without explicit 
dynamics \cite{Rau2016, Udagawa2016, Mizoguchi2017}.
Here, the same basic mechanism, of pinch points 
``bent up'' in energy, presumably applies. 
However, alternative methods would be needed to elaborate on how this happens.
Another important open problem is the fate of half moons as a fully entangled 
quantum spin liquid (QSL).
In this context, it is interesting to ask how the pinch points and half moons of 
the proximate spin liquid \NZO\ would evolve, if it were possible to close the gap to 
the flat band of excitations carrying pinch points \cite{benton16-PRB94,Petit2016}.
In this case we anticipate that the flat band will evolve into the photons 
of a quantum spin ice \cite{benton16-PRB94}, while the dispersing band carrying 
the half moons must connect with its topological, 
``magnetic monopole'' excitations \cite{huang18-PRL120}.  
Explicit calculations for QSL's, however, remain to be carried out.
Finally, while completing this work, we learned of a parallel study 
by Mizoguchi {\it et al.}, which reports complementary results 
for a different model \cite{ludo-private-communication}.


\indent
{\it Acknowledgements. }
This work was supported by the Theory of Quantum Matter Unit, Okinawa Institute 
of Science and Technology Graduate University (OIST).
H.Y. is supported by 
Japan Society for the Promotion of Science (JSPS) Research 
Fellowship for Young Scientists. 
The authors would like to thank Owen Benton, Bella Lake, Mathieu Taillefumier, 
and Alexandra Turrini for helpful discussions.
Numerical calculations are carried out using HPC Facilities provided by OIST.
H.Y. acknowledges the hospitality of the Yukawa Institute for Theoretical Physics at Kyoto University,
where part of this work was  carried out during the workshop 
``Novel Quantum States in Condensed Matter 2017" (NQS2017,  YITP-T-17-01).


\bibliography{paper.bib}

\clearpage
\widetext


\section*{Supplementary material}

# Supplemental Materials for Half moons are pinch points with dispersion

Han Yan (闫寒),[1, *] Rico Pohle,[1] and Nic Shannon[1]

[1]*Okinawa Institute of Science and Technology Graduate University, Onna-son, Okinawa 904-0412, Japan*

(Dated: October 12, 2018)

## I. KAGOME LATTICE MODEL AND ASSOCIATED EQUATIONS OF MOTION

As the simplest representative case, exhibiting both pinch points and half moons, we consider a Heisenberg model on the Kagome lattice

$$\mathcal{H} = J \sum_{\langle ij \rangle} \mathbf{S}_i \cdot \mathbf{S}_j - H \sum_i S_i^z, \tag{1}$$

where $H > 0$ is the applied magnetic field, $J > 0$ is an antiferromagnetic exchange interaction, and the sum $\langle ij \rangle$ runs over the nearest-neighbor bonds of the lattice, shown in Fig. 1. For magnetic field $H > H_c = 6J$, the ground state of this model is a fully-saturated state, in which all spins are aligned with magnetic field. In this limit one-magnon excitations are exact eigenstates, which can be described exactly using either linear spin wave theory (LSW), or the equations of motion we derive below. And since both pinch points and half moons can be observed in the one-magnon bands, this is a sufficiently general case to explore the connection between them.

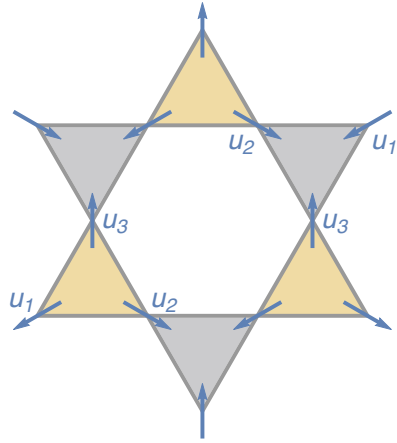

FIG. 1. Kagome lattice associated with the Heisenberg model $\mathcal{H}$ [Eq. (1)], and associated vectors $\mathbf{u}_i$ [Eq. (5)], used in decomposition of fields [Eq. (4)].

We start by considering the equation of motion (EoM) for spins on a lattice. The Heisenberg EoM for a single spin $\mathbf{S}_i = (S_i^x, S_i^y, S_i^z)$ are given by

$$\frac{d\mathbf{S}_i}{dt} = \frac{i}{\hbar}[\mathcal{H}, \mathbf{S}_i], \tag{2}$$

where, in what follows, we set $\hbar = 1$. The primitive unit cell for the Kagome lattice contains three spins, denoted $\mathbf{S}_i, i = 1, 2, 3$. For $H \gg J$, the ground state is saturated as $\langle S_i^z \rangle = 1$, $\langle S_i^{x,y} \rangle = 0$. At linear order, there are six degrees of freedom in each unit cell, namely $S_i^x, S_i^y$ for $i = 1, 2, 3$. Their EoM's are

$$\partial_t S_i^x = J \sum_j (S_j^y - S_i^y) + S_i^y H,$$
$$\partial_t S_i^y = -J \sum_j (S_j^x - S_i^x) - S_i^x H. \tag{3}$$

---

* han.yan@oist.jp

where the sum upon $j$ runs over the nearest-neighbour bonds of the site $i$.

## II. EQUATIONS OF MOTION FOR CONTINUOUS FIELDS

The six degrees of freedom $S_i^x$, $S_i^y$ for $i = 1, 2, 3$ in each unit cell can be reorganized as irreducible representations of the lattice symmetry:

$$\mathbf{m}_x = \sum_{i=1}^{3} S_i^x \mathbf{u}_i, \quad \xi_x = \sum_{i=1}^{3} S_i^x,$$
$$\mathbf{m}_y = \sum_{i=1}^{3} S_i^y \mathbf{u}_i, \quad \xi_y = \sum_{i=1}^{3} S_i^y, \tag{4}$$

where

$$\mathbf{u}_1 = (-\sqrt{3}/2, -1/2), \quad \mathbf{u}_2 = (\sqrt{3}/2, -1/2), \quad \mathbf{u}_3 = (0, 1), \tag{5}$$

are local unit vectors to help us build the the irreducible representations [cf Fig 1]. Note here $x, y$ in the subscript denote different fields, not their spatial components. The explicit expressions for $\mathbf{m}_x$ and $\mathbf{m}_y$ are

$$\mathbf{m}_x = (\sqrt{3}/2(-S_1^x + S_2^x), -S_1^x/2 - S_2^x/2 + S_3^x),$$
$$\mathbf{m}_y = (\sqrt{3}/2(-S_1^y + S_2^y), -S_1^y/2 - S_2^y/2 + S_3^y). \tag{6}$$

We can rewrite the discrete, microscopic EoMs [Eq. 3] in the continuous limit in terms of $\mathbf{m}_\nu$ and $\xi_\nu$:

$$\partial_t \mathbf{m}_x = \frac{J}{4} \nabla (\nabla \cdot \mathbf{m}_y) + \tilde{H} \mathbf{m}_y,$$
$$\partial_t \mathbf{m}_y = -\frac{J}{4} \nabla (\nabla \cdot \mathbf{m}_x) - \tilde{H} \mathbf{m}_x,$$
$$\partial_t \xi_x = -\frac{J}{4} \nabla^2 \xi_y + H \xi_y,$$
$$\partial_t \xi_y = \frac{J}{4} \nabla^2 \xi_x - H \xi_x, \tag{7}$$

where $\tilde{H} = H - 6J$. We note that the definitions of divergence and curl in two dimension, shown in Table I, are slightly different from those in three dimensions.

TABLE I. Summary of definitions of divergence/gradient and curl/scalar curl in 2D.

|  | Notation | Definition | Acting on vector field $\mathbf{v}$ | Acting on scalar field $\phi$ |
| --- | --- | --- | --- | --- |
| Divergence/gradient | $\nabla$ | $(\partial_x, \partial_y)$ | $\partial_x v_x + \partial_y v_y$ | $(\partial_x \phi, \partial_y \phi)$ |
| Curl/scalar curl | $\nabla_\perp$ | $(-\partial_y, \partial_x)$ | $-\partial_y v_x + \partial_x v_y$ | $(-\partial_y \phi, \partial_x \phi)$ |

Except for the constant energy shift terms with coefficient $\tilde{H}$ or $H$, all terms on the right hand side of the EoMs are of the form $\nabla$ acting on a scalar or vector field. This property of the EoMs suggests that we can decompose $\mathbf{m}_\nu$ into $\mathbf{m}_\nu^{\text{curl}}$ and $\mathbf{m}_\nu^{\text{div}}$, by standard Helmholtz-Hodge decomposition [1],

$$\mathbf{m}_\nu = \mathbf{m}_\nu^{\text{curl}} + \mathbf{m}_\nu^{\text{div}}, \tag{8}$$

where

$$\mathbf{m}_\nu^{\text{curl}} = \nabla_\perp a_\nu, \quad \mathbf{m}_\nu^{\text{div}} = \nabla \phi_\nu, \tag{9}$$

and $a_\nu, \phi_\nu$ are scalar potentials. We note that, in making the decomposition Eq. (8), we have assumed vanishing harmonic components of the field $\mathbf{m}_\nu$. Written in this form, the vector field $\mathbf{m}_\nu$ has two components, which are characterised by the two fields $a_\nu$ and $\phi_\nu$. And by analogy with magnetostatics, $a_\nu$ plays the role of a gauge field, while $\phi_\nu$ corresponds to the potential coming from a magnetic charge.

Exploiting the fact that

$$\nabla_\perp \cdot \mathbf{m}_\nu^{\text{div}} = 0, \quad \nabla \cdot \mathbf{m}_\nu^{\text{curl}} = 0, \tag{10}$$

we can get simplified EoMs, in which the 6 fields $\mathbf{m}_\nu^{\text{curl}}$, $\mathbf{m}_\nu^{\text{div}}$ and $\xi_\nu$ all decouple from one other to yield

$$\partial_t \mathbf{m}_x^{\text{curl}} = +\tilde{H}\mathbf{m}_y^{\text{curl}}, \quad \partial_t \mathbf{m}_y^{\text{curl}} = -\tilde{H}\mathbf{m}_x^{\text{curl}}, \tag{11}$$

$$\partial_t \mathbf{m}_x^{\text{div}} = \frac{J}{4}\nabla(\nabla \cdot \mathbf{m}_y^{\text{div}}) + \tilde{H}\mathbf{m}_y^{\text{div}}, \quad \partial_t \mathbf{m}_y^{\text{div}} = -\frac{J}{4}\nabla(\nabla \cdot \mathbf{m}_x^{\text{div}}) - \tilde{H}\mathbf{m}_x^{\text{div}}, \tag{12}$$

$$\partial_t \xi_x = -\frac{J}{4}\nabla^2 \xi_y + H\xi_y, \quad \partial_t \xi_y = \frac{J}{4}\nabla^2 \xi_x - H\xi_x. \tag{13}$$

Conventional pinch points are encoded in the field $\mathbf{m}_\nu^{\text{curl}}$. For the purpose of understanding these we can use Eq. (9) to rewrite Eq. (11) in terms of $a_\nu$. This leads to the result

$$\partial_t a_x = \omega_0 a_y, \quad \partial_t a_y = -\omega_0 a_x, \tag{14}$$

quoted in the Main Text. Meanwhile, "half moons" are encoded in the field $\mathbf{m}_\nu^{\text{div}}$. And, for the purpose of understanding these, we can use Eq. (9) to rewrite Eq. (12) in terms of $a_\nu$ and $\phi_\nu$. This leads to the result

$$\partial_t \phi_x = \left[\omega_0 + D\nabla^2\right]\phi_y, \quad \partial_t \phi_y = \left[-\omega_0 - D\nabla^2\right]\phi_x, \tag{15}$$

quoted in the Main Text.

## III. ANALYSIS OF EQUATIONS OF MOTION

### A. Dispersion relations

Firstly we examine the dispersion relations which follow from these EoM. In the field-theory approach, the dispersion relations can be obtained by Fourier transforming EoMs [cf Eq. (11), Eq. (12)] into energy-momentum space $(\mathbf{q}, \omega)$, and solving for the eigenvalues $\omega$ as a function of $\mathbf{q}$. This yields

$$\omega_{\text{top}}(\mathbf{q}) = H - \frac{J}{4}\mathbf{q}^2, \tag{16}$$

$$\omega_{\text{mid}}(\mathbf{q}) = \omega_0 + \frac{J}{4}\mathbf{q}^2, \tag{17}$$

$$\omega_{\text{bot}}(\mathbf{q}) = \omega_0, \tag{18}$$

where

$$\omega_0 = \tilde{H} = H - 6J. \tag{19}$$

As discussed in Section IV, equivalent results for the dispersion at long wavelength can also be derived within a standard linear spin-wave calculation.

### B. Singular structures in scatterring

We now use the field theory to examine the contribution to the equal-time structure factor coming from each band of spin excitations. We consider first the structure factor associated with

$$S_{x,x}^{\text{curl}}(\mathbf{q}) \equiv \langle m_{x,x}^{\text{curl}}(\mathbf{q}) m_{x,x}^{\text{curl}}(-\mathbf{q})\rangle. \tag{20}$$

The field theory approach developed above leads to the prediction

$$\langle m_{\nu,\alpha}^{\text{curl}}(\mathbf{q},\omega) m_{\nu',\beta}^{\text{curl}}(-\mathbf{q},,\omega)\rangle \propto \delta_{\nu\nu'}\delta(\omega - \omega_0)(1 - q_\alpha q_\beta/\mathbf{q}^2), \tag{21}$$

where $\omega_0 = H - 6J$, $\alpha, \beta = x, y$ are the spatial components of vector field $\mathbf{m}_\nu^{\text{curl}}$ and $\mathbf{m}_{\nu'}^{\text{curl}}$. Fixing $\alpha = \beta = x = \nu = \nu' = x$, and integrating over $\omega$, we find

$$S_{x,x}^{\text{curl}}(\mathbf{q}) = \langle m_{x,x}^{\text{curl}}(\mathbf{q}) m_{x,x}^{\text{curl}}(-\mathbf{q})\rangle \propto 1 - \frac{q_x^2}{\mathbf{q}^2}, \tag{22}$$

i.e. a pinch point. As discussed in Section IV, equivalent results can also be derived within a standard linear spin-wave (LSW) calculation.

We now consider the energy-integrated (equal-time) structure factor characterising the dispersing band associated with half moons, i.e.

$$S_{x,x}^{\mathsf{div}}(\mathbf{q}) \equiv \langle m_{x,x}^{\mathsf{div}}(\mathbf{q}) m_{x,x}^{\mathsf{div}}(-\mathbf{q}) \rangle. \tag{23}$$

An analysis similar to that leading to Eq.(22), yields the result quote for the assocaited dynamical structure factor in Eq. (12) of the main text, i.e.,

$$\langle m_{\nu,\alpha}^{\mathsf{div}}(\mathbf{q},\omega) m_{\nu',\beta}^{\mathsf{div}}(-\mathbf{q},\omega) \rangle \propto \delta_{\nu\nu'} \delta(\omega - \omega_0 - D\mathbf{q}^2) \times (1 - \tilde{q}_\alpha \tilde{q}_\beta / \mathbf{q}^2),$$

where $D = J/4$. It follows that the field-theory prediction for the energy-integrated structure factor $S_{x,x}^{\mathsf{div}}(\mathbf{q})$ also has the structure of a pinch point

$$S_{x,x}^{\mathsf{div}}(\mathbf{q}) = \langle m_{x,x}^{\mathsf{div}}(\mathbf{q}) m_{x,x}^{\mathsf{div}}(-\mathbf{q}) \rangle \propto 1 - \frac{q_y^2}{\mathbf{q}^2}, \tag{24}$$

but one perpendicular to the pinch point in Eq. (22). As discussed in Section IV, equivalent results can also be derived within a standard linear spin-wave (LSW) calculation.

## IV. COMPARISON BETWEEN FIELD THEORY AND LINEAR SPIN WAVE THEORY

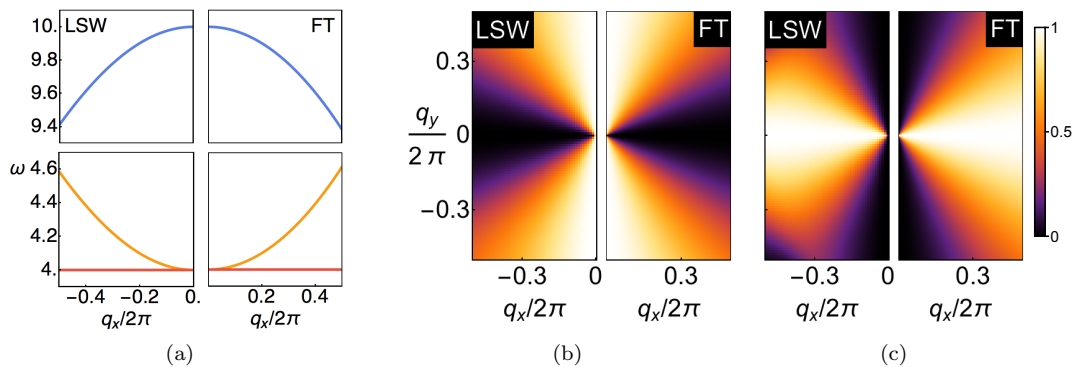

FIG. 2. Comparison of dynamical correlation functions calculated from linear spin wave (LSW) theory and field theory (FT). Results from LSW and FT fit well for a fairly large region near the Γ-point. The calculation is done for parameter $J = 1.0$ and $H = 10.0$. (a) Dispersion of spin exciations along $(q_x, 0)$ near the Γ-point, calculated by LSW (left panel) and field theory (right panel). (b) Equal-time structure factor $S_{x,x}^{\mathsf{curl}}(\mathbf{q})$ [Eq. (20)] calculated within LSW (left panel) and field theory (right panel), showing pinch points associated with the flat band of excitations. (c) Equal-time structure factor $S_{x,x}^{\mathsf{div}}(\mathbf{q})$ [Eq. (23)] calculated within LSW (left panel) and field theory (right panel), showing pinch points associated with the dispersing band of excitations.

To validate this approach, it is helpful to compare the predictions of the field theory, quoted above, with equivalent results calculated within standard linear spin-wave theory (LSW). We first consider the dispersion relations for the three bands in the vicinity of the Γ-point. The results are summarized in Fig 2a; good agreement is achieved for a wide range of **q** centered on the Γ-point.

We can also compare predictions for dynamical structure factors. As as a representative case, we can evaluate e.g.

$$\langle m_{x,x}(\omega, \mathbf{q}) m_{x,x}(\omega, -\mathbf{q}) \rangle = \langle \frac{\sqrt{3}}{2}(S_2^x - S_3^x)(\mathbf{q}) \frac{\sqrt{3}}{2}(S_2^x - S_3^x)(-\mathbf{q}) \rangle, \tag{25}$$

using the LSW eigenvectors for the band $\mathbf{m}_\nu^{\mathsf{div}}$ (the dispersing band in the middle of Fig 2(a)). The associated predictions for the equal-time structure factors $S_{x,x}^{\mathsf{curl}}(\mathbf{q})$ and $S_{x,x}^{\mathsf{div}}(\mathbf{q})$, calculated within the two approaches, are shown in Fig 2(b) and Fig 2(c). Once again, the agreement is excellent for a large range of **q** centered on the Γ-point.

## V. EMERGENT GAUGE DESCRIPTION IN 3D AND 2D COULOMBIC PHASES

The exotic properties of Coulombic phases, such as the spin liquid found in spin ice, follow from the local degeneracy of their ground state manifold, and are most easily understood by introducing an effective action which is invariant under a set of local "gauge" transformations. This "emergent gauge invariance" is a deep and subtle concept, which has much in common with the $U(1)$ gauge invariance present in classical electromagnetism.

In what follows we discuss the connection between the "Coulombic" aspects of the models we study, as manifest in their pinch points (and half moons), and an emergent $U(1)$ gauge description. In particular, we explore what it means for such an emergent invariance to be "broken", by the introduction of terms in the Hamiltonian which lift the ground-state degeneracy. We begin by reviewing the ways in which local gauge symmetries enter treatments of three-dimensional pyrochlore spin ice, and two-dimensional square ice. An equivalent discussion of the connection between an emergent gauge theory and pinch points can be found in two articles on Coulomb phases by Henley [2, 3].

The ground states of spin ice are determined by a local constraint, the "ice rule", which specifies that two spins point into, and two spins point out of, each tetrahedron on the lattice. Remembering that each of these spins contributes to the magnetisation of the system, we can write this local constraint as

$$\nabla \cdot \mathbf{m} = 0 \tag{26}$$

where $\mathbf{m}$ is a continuum field describing the magnetisation at a suitably course-grained scale [2]. The divergence free property motivates us to treat $\mathbf{m}$ as a magnetic field, and make the analogy between these classical spin systems and magnetostatics. In the case of three-dimensional spin ice, we can resolve the constraint Eq. (26) as

$$\mathbf{m} = \nabla \times \mathbf{A}, \tag{27}$$

where the three-dimensional vector field $\mathbf{A}$ possesses an "emergent" gauge symmetry

$$\mathbf{A} \to \mathbf{A} + \nabla \lambda. \tag{28}$$

It is important to note that, in classical spin ice, just as in classical electromagnetism, the "gauge" field $\mathbf{A}$ [Eq. (28)] does not have have a direct physical interpretation; this is reserved for the "magnetic" field $\mathbf{m}$ [Eq. (26)]. None the less, the gauge transformation, Eq. (28), is a symmetry of the effective action which describes the low-energy properties of spin ice [2], and underlies all of its exotic properties. And in quantum models of spin ice, just as in quantum electrodynamics, the emergent gauge field $\mathbf{A}$ has direct physical consequences, which include emergent photon excitations, that can be observed in experiment [4–6]. For this reason, quantum spin ice is usually regarded as physical realisation of a compact, frustrated, $U(1)$ lattice gauge theory [7].

The same "ice rules" constraint, Eq. (26), applies in the two-dimensional analogue of spin ice, usually referred to as "square ice". However the fact that the magnetization $\mathbf{m}$ is a two-dimensional vector, changes the way in which gauge symmetry enters the problem. In this case, we can resolve the constraint as

$$\mathbf{m} = \nabla_\perp a \tag{29}$$

where $a$ is a scalar and $\nabla_\perp$ is defined in Table I. In this case, however, the identification between $\mathbf{m}$ and $a$ is unique, so there is no room for gauge redundancy [2]. None the less, we can still construct a Helmholtz decomposition of $\mathbf{m}$, and the absence of an (emergent) gauge symmetry in the two-dimensional classial effective action, does not change any of the arguments we develop.

What matters instead is that any two configurations of gauge field $\mathbf{A}, \mathbf{A}'$, or $a$, $a'$ that are *physically* different,

$$\mathbf{m} = \nabla \times \mathbf{A} \neq \mathbf{m}' = \nabla \times \mathbf{A}' \qquad \text{for spin ice,} \tag{30}$$

$$\mathbf{m} = \nabla_\perp a \neq \mathbf{m}' = \nabla_\perp a' \qquad \text{for square ice,} \tag{31}$$

remain degenerate ground states, for which both $\mathbf{m}$ and $\mathbf{m}'$ satisfy Eq. (26). This means that the effective action describing a Coulombic phase - i.e. one leading to pinch points - cannot contain any terms which lift this ground-state degeneracy, either (1) by generating a finite energy gap to excitaitions, or (2) by distinguishing between the energy of $\mathbf{A}$ from $\mathbf{A}'$ (equivalently, $a$ from $a'$).

The different ways in which this degeneracy can be lifted can be described within a continuum field theory through either a "mass" term in the effective action

$$\Delta \mathcal{H}_{\text{mass-spin ice}} = m^2 \mathbf{A}^2, \quad \Delta \mathcal{H}_{\text{mass-square ice}} = m^2 a^2, \tag{32}$$

which generates an energy gap to all excitations [Case (1)] or a term with spatial derivative,

$$\Delta \mathcal{H}_{\text{prop-spin ice}} = (\nabla \cdot \mathbf{A})^2, \quad \Delta \mathcal{H}_{\text{prop-square ice}} = (\nabla a)^2. \tag{33}$$

which distinguishes the energy of fields with different wave vector [Case (2)]. Physically, both terms lift the degeneracy of the ground-state manifold, so that spin configurations corresponding to *physically* different $\mathbf{A}(\mathbf{r})$ are nolonger degenerate. In the problems we consider, such terms are not generated spontaneously, but occur through terms in the Hamiltonian which explicitly lift the ground state degeneracy associated with a Coulomb phase.

It should be noted that this route to breaking an emergent gauge symmetry is fundamentally different from the Higgs mechanism, which can also introduce a mass to the gauge boson. In the Higgs mechanism, gauge-symmetry remains unbroken, while a *global* symmetry is broken instead. Here the terms Eq. (32, 32) *explicitly* violate the gauge invariance of the (effective) action. Terms of this form cannot be generated spontaneously by fluctuations, since for a local invariance this is forbidden by Elitzur's theorem. However, they can easily be introduced by changing the parameters of a microscopic model, for example by introducing a magnetic field which couples to the magnetisation of the spins.

In this work, we consider the Kagome-lattice antiferromagnet [Eq. (1)] as the prototype of both a classical spin liquid [8–10], where pinch points can be linked to an emergent gauge field [2], and of a system displaying half-moon features in its dynamics [11, 12]. We study this model primarily in the high-field limit $H > H_c = 6J$, finding both a flat band possessing pinch points, and two dispersing bands exhibiting half moons, above a gap which closes for $H \to H_c$. The presence of the gap at

$$\omega = E_0 \tag{34}$$

is a consequence of terms of the form Eq. (32) turned on in the effective action describing the field-saturated state, while the flatness of the band with pinch points implies the absence of terms of the form Eq. (33). So we have to introduce a phenomenological term

$$\mathcal{H}_{\mathsf{mass}} = m^2 a^2, \quad m = E_0. \tag{35}$$

The existence of the pinch points follows from the divergence-free condition [Eq. (26)] [2].

Finally, when the proper quantum fluctuation is introduced, the system is allowed to tunnel between different classical ground state configurations. Essentially it introduces a dynamical term for the gauge fields,

$$\mathcal{H}_{\mathsf{dy\text{-}spin\ ice}} = (\dot{\mathbf{A}})^2, \quad \mathcal{H}_{\mathsf{dy\text{-}square\ ice}} = (\dot{a})^2, \tag{36}$$

which plays the role of electric fields. Such quantum tunneling terms will upgrade the classical spin liquid into a quantum spin liquid, and recovers the full electromagnetism [4–6].

## VI. APPLICATION TO $\mathrm{Nd_2Zr_2O_7}$

The field theory developed in the Main Text is sufficient to reproduce inelastic neutron scattering results of $\mathrm{Nd_2Zr_2O_7}$. $\mathrm{Nd_2Zr_2O_7}$ is a is a Pyrochlore material which exhibits "momentum fragmentation", realising a ground state with "all-in all-out" (AIAO) magnetic order, with a dynamical spin liquid encoded in its excitations [13, 14]. Following [14], its effective Hamiltonian can be written as

$$\mathcal{H}_{\mathsf{NZO}} = \sum_{\langle ij \rangle} [\tilde{J}_x \tilde{\tau}_i^{\tilde{x}} \tilde{\tau}_j^{\tilde{x}} + \tilde{J}_y \tilde{\tau}_i^{\tilde{y}} \tilde{\tau}_j^{\tilde{y}} + \tilde{J}_z \tilde{\tau}_i^{\tilde{z}} \tilde{\tau}_j^{\tilde{z}}], \tag{37}$$

where $\tilde{\tau}_i^{\tilde{\alpha}}$ are the spin-half operators with a global rotation. Up to a constant coefficient, the magnetic moment on lattice site $i$ is given by

$$\mathbf{m}_i = \cos\theta \mathbf{m}_i^{\tilde{z}} + \sin\theta \mathbf{m}_i^{\tilde{x}}, \tag{38}$$

where $\mathbf{m}_i^{\tilde{\alpha}}$ is defined as

$$\mathbf{m}_i^{\tilde{\alpha}} = \tilde{\tau}_i^{\tilde{\alpha}} \hat{\mathbf{z}}_i. \tag{39}$$

The values of the parameters

$$\tilde{J}_x = 0.103 meV, \quad \tilde{J}_y = 0.0 meV, \quad \tilde{J}_z = -0.047 meV, \quad \theta = 0.83, \tag{40}$$

are taken from [14].

Our aim is to reproduce the spin-spin correlation measured by neutron scattering

$$S^{\text{theory}}(\mathbf{q},\omega) = \cos^2\theta S^{\tilde{z}}(\mathbf{q},\omega) + \sin^2\theta S^{\tilde{x}}(\mathbf{q},\omega), \tag{41}$$

where

$$S^{\tilde{\alpha}}(\mathbf{q},\omega) = \int dt e^{-i\omega t}(\delta_{\beta\gamma} - \frac{q_\beta q_\gamma}{q^2})\langle m^{\tilde{\alpha}}_\beta(\mathbf{q})m^{\tilde{\alpha}}_\gamma(-\mathbf{q})\rangle. \tag{42}$$

Here $\tilde{\alpha} = \tilde{z}, \tilde{x}$ denote different vector fields, and $\beta, \gamma = x, y, z$ denote the spatial components.

The essential physics of fragmentation is that $\mathbf{m}^{\tilde{z}}_i$ forms the AIAO order, and it pushes the $\mathbf{m}^{\tilde{x}}_i$ fluctuation to finite energy. Furthermore, the fluctuation of $\mathbf{m}^{\tilde{x}}_i$ obeying $\nabla\cdot\mathbf{m}^{\tilde{x}}_i = 0$ decouples dynamically, and forms a flat band. The divergence-free condition gives rise to pinch points on the band. This is exactly the physics of $\mathbf{m}^{\text{curl}}_\nu$ in the Main Text.

From the Main Text we also know that the divergence-full part $\nabla\times\mathbf{m}^{\tilde{x}}_i = 0$ of its fluctuation will form another band that is curved, and encodes the perpendicular pinch points as half moons. In this case the system is in three-dimensional, but that does not make any qualitative difference.

Therefore, we can conclude that $S^{\tilde{z}}(\mathbf{q},\omega)$ contains the Bragg peaks of the ordered $\mathbf{m}^{\tilde{z}}_i$. We can ignore them. $S^{\tilde{x}}(\mathbf{q},\omega)$ has the structure of the spin-spin correlation we discussed in the Main Text. It contains a flat band with pinch points, and a curved band with half moons.

Let us now compute $S^{\tilde{x}}(\mathbf{q},\omega)$ quantitatively. Applying the Helmholtz decomposition [cf. Ref.[14]], the vector fields $\mathbf{m}^{\tilde{\alpha}}, \alpha = x, y$, can be written as

$$\mathbf{m}^{\tilde{\alpha}} = \mathbf{m}^{\tilde{\alpha},\text{div}} + \mathbf{m}^{\tilde{\alpha},\text{curl}} = \nabla a^{\tilde{\alpha}} + \nabla\times\mathbf{A}^{\tilde{\alpha}}. \tag{43}$$

Their equations of motion are [14]:

$$\partial_t \mathbf{A}^{\tilde{x}} = -(3|\tilde{J}_z| - 1\tilde{J}_y)\mathbf{A}^{\tilde{y}}, \tag{44}$$

$$\partial_t \mathbf{A}^{\tilde{y}} = (3|\tilde{J}_z| - 1\tilde{J}_x)\mathbf{A}^{\tilde{x}}, \tag{45}$$

$$\partial_t a^{\tilde{x}} = -(\frac{\tilde{J}_y}{2}\nabla^2 + (3|\tilde{J}_z| - 1\tilde{J}_y))a^{\tilde{y}}, \tag{46}$$

$$\partial_t a^{\tilde{y}} = (\frac{\tilde{J}_x}{2}\nabla^2 + (3|\tilde{J}_z| - 1\tilde{J}_x))a^{\tilde{x}}. \tag{47}$$

From Eqs. (44),(45), we know that $\mathbf{m}^{\tilde{\alpha},\text{curl}}$ form a flat band with pinch points at energy

$$\omega_1 = \left[(3|\tilde{J}_z| - \tilde{J}_x)(3|\tilde{J}_z| - \tilde{J}_y)\right]^{1/2}. \tag{48}$$

From Eqs. (46),(47), we know that $\mathbf{m}^{\tilde{\alpha},\text{div}}$ form a curved band with half moons. Its dispersion relation is

$$\omega_2(\mathbf{q}) = \left[(\tilde{J}_x\mathbf{q}^2/2 + 3|\tilde{J}_z| - \tilde{J}_x)(\tilde{J}_y\mathbf{q}^2/2 + 3|\tilde{J}_z| - \tilde{J}_y)\right]^{1/2}$$
$$\approx [(3|\tilde{J}_z| - \tilde{J}_x)(3|\tilde{J}_z| - \tilde{J}_y)]^{1/2}\times$$
$$\left[1 + \frac{\tilde{J}_x(3|\tilde{J}_z| - \tilde{J}_y) + \tilde{J}_y(3|\tilde{J}_z| - \tilde{J}_x)}{4(3|\tilde{J}_z| - \tilde{J}_x)(3|\tilde{J}_z| - \tilde{J}_y)}\mathbf{q}^2\right]. \tag{49}$$

Combining the band dispersion relations Eqs.(48),(48), and the knowledge of the spectral weight distributions on these bands from the Main Text Eq. (8,12), one can write down $S^{\tilde{x}}(\mathbf{q},\omega)$ entirely. For $\mathbf{q} = (0,0,2) + (\delta h, \delta h, \delta l)$, we have

$$S^{\text{theory}}(\mathbf{q},\omega) = sin^2\theta S^{\tilde{x}}(\mathbf{q},\omega) = \delta(\omega - \omega_1)\times sin^2\theta(1 + \frac{(\delta h)^2}{(\delta h)^2 + (\delta l)^2}) + \delta(\omega - \omega_2(\mathbf{q}))\times sin^2\theta\frac{(\delta l)^2}{(\delta h)^2 + (\delta l)^2}. \tag{50}$$

Finally let us fit the theoretical results with experiments. The major issues are that in experiment there is a strong incoherent background signal, and the experimental resolution is finite. For the first issue we introduce a Gaussian background

$$I^{\text{bkg}}(\mathbf{q},\omega) = A\exp(-\frac{\omega^2}{\sigma^2}). \tag{51}$$

For the second issue we introduce a Gaussian convolution of $S^{\text{theory}}(\mathbf{q},\omega)$ with the same width $\sigma$

$$I^{\text{theory}}(\mathbf{q},\omega) = \int d\omega' B \exp(-\frac{(\omega-\omega')^2}{\sigma^2}) \times S^{\text{theory}}(\mathbf{q},\omega'), \qquad (52)$$

and use the sum of the two

$$I^{\text{exp}}(\mathbf{q},\omega) = I^{\text{bkg}}(\mathbf{q},\omega) + I^{\text{theory}}(\mathbf{q},\omega) \qquad (53)$$

to represent the experimental results.

There are three fitting parameters $A, B, \sigma$ in $I^{\text{exp}}(\mathbf{q},\omega)$. To determine them, we visually fit it against the red curve in Fig. 4(b) of [13], and obtain

$$A = 1.0 meV, \quad B = 0.24, \quad \sigma = 0.027 meV. \qquad (54)$$

There is no freely adjustable parameters left. $I^{\text{exp}}(\mathbf{q} = (0.9, 0.9, 0.9), \omega)$ as a function of $\omega$ is plotted in Fig. 3, to be compared to Fig. 4(b) of [13].

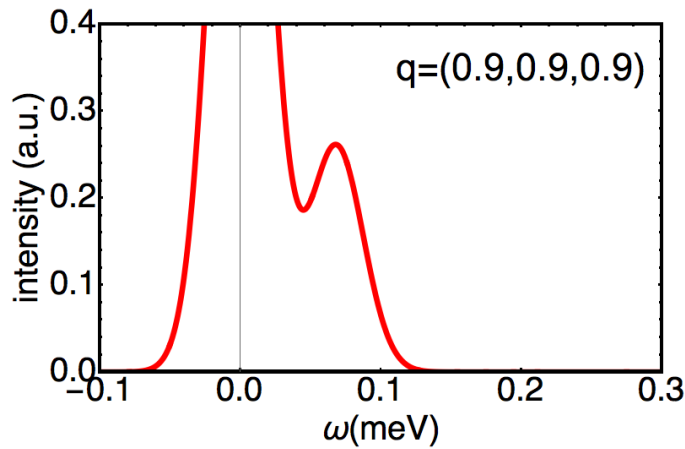

FIG. 3. Signal strength $I^{\text{exp}}(\mathbf{q},\omega)$ at $\mathbf{q} = (0.9, 0.9, 0.9)$ computed by the field theory with proper Gaussian evolutions and background [cf Eq. (53)], fitted against the red curve of Fig. 4(b) of [13].

With the full expression of $I^{\text{exp}}(\mathbf{q},\omega)$ given, we proceed to plot the inelastic neutron scattering results, which are shown in Fig. 4 of the Main Text. They are in good agreement of Fig. S3 of the Supplemental Information of [13], for regions close to the $\Gamma$-points.

## VII. NUMERICAL SIMULATION

### A. Classical Monte Carlo Simulations

Numerical simulations have been performed for classical vector spins of length $|\mathbf{S}| = 1$, by evaluating Eq. (1) for $H = 0$. Monte Carlo (MC) simulations were performed by using the local heat-bath algorithm [15, 16], in combination with parallel tempering [17, 18], and over-relaxation [19] on systems with 24300 spins. A single hybrid MC step consists of $N$ local heat-bath updates on randomly chosen sites, followed by two over-relaxation steps for the whole lattice, each comprising a $\pi$-rotation of each spin about their local exchange field. Simulations were performed in parallel for replicas at 400 different temperatures, with replica-exchange initiated by the parallel tempering algorithm every 100th MC step. Thermodynamic quantities were averaged over $10^6$ statistically independent samples, after simulated annealing ($10^6$ MC steps) and thermalisation ($10^6$ MC steps). In addition, all measured quantities have been averaged over 20 independent simulations.

## B. Molecular Dynamics

Molecular dynamics (MD) simulations are based on the numerical integration of the Heisenberg equations of motion

$$\frac{d\mathbf{S}_i}{dt} = \frac{i}{\hbar}\big[\mathcal{H}, \mathbf{S}_i\big] = J \sum_j \mathbf{S}_j \times \mathbf{S}_i, \tag{55}$$

where $j$ accounts for all nearest-neighbouring sites of $i$. Spin configurations for MD simulations were taken from a thermal ensemble of spins generated by classical MC simulations of $\mathcal{H}$ [Eq. (1)], for temperatures as indicated in Fig. 3 (Main Text). Numerical integration of Eq. (55) has carried out using a $4^{th}$ order Runge-Kutta algorithm, as described in [20, 21].

The dynamical structure factor

$$S(\mathbf{q}, \omega) = \frac{1}{\sqrt{N_t} N} \sum_{i,j}^{N} e^{i\mathbf{q}(\mathbf{r}_i - \mathbf{r}_j)} \sum_{n}^{N_t} e^{i\omega \; n\delta t} \langle \mathbf{S}_i(0) \cdot \mathbf{S}_j(t) \rangle, \tag{56}$$

was calculated by using a Fast Fourier Transform (FFT) [22], and has been averaged over spin dynamics obtained from 1000 independent initial spin configurations. MD simulations shown in Fig. 3 (Main Text) were performed for $N_t = 1000$ time steps and maximally resolvable frequency limit of $\omega_{\max} = 6J$. Each resolving time-increment $\delta t$ can be obtained by

$$\delta t = \frac{t_{\max}}{N_t} = \frac{2\pi}{\omega_{\max}} \; . \tag{57}$$

Additionally, the time sequence of spin configurations has been multiplied by a Gaussian envelop, in order to avoid numerical artefacts, like the Gibbs phenomenon [1] coming from discontinuities of the finite time-window at $t = 0$ and $t = t_{\max}$. This imposes a Gaussian energy convolution on the numerically-obtained $S(\mathbf{q}, \omega)$ in Fig. 3 (Main Text) of FWHM $\approx 0.02J$.

---


\end{document}